\documentclass[aps,prl,twocolumn,floatfix]{revtex4-1}
\usepackage{graphicx}
\usepackage{natbib}

\newlength{\onecolfig}
\newlength{\twocolfig}
\newcommand{\ion}[2]{\mbox{$^{#2}$#1$^+$}}
\newcommand{\Ca}[1]{\ion{Ca}{#1}}

\newcommand{\Yb}[1]{\ion{Yb}{#1}}

\newcommand{\hfslev}[3]{\mbox{#1$^{\mbox{\tiny$#3$}}_{\mbox{\tiny$#2$}}$}}

\newcommand{\unit}[1]{\,\mbox{#1}}
\newcommand{\mHz}{\unit{mHz}}
\newcommand{\Hz}{\unit{Hz}}
\newcommand{\kHz}{\unit{kHz}}
\newcommand{\MHz}{\unit{MHz}}
\newcommand{\GHz}{\unit{GHz}}

\newcommand{\rad}{\unit{rad}}

\newcommand{\urad}{\unit{$\mu$rad}}

\newcommand{\um}{\unit{$\mu$m}}

\newcommand{\s}{\unit{s}}

\renewcommand{\min}{\unit{min}}

\newcommand{\ms}{\unit{ms}}
\newcommand{\us}{\unit{$\mu$s}}

\newcommand{\degree}{\mbox{$^{\circ}$}}

\newcommand{\G}{\unit{G}}
\newcommand{\mG}{\unit{mG}}
\newcommand{\uG}{\unit{$\mu$G}}

\newcommand{\amp}{\unit{A}}

\newcommand{\ish}{\mbox{$\sim$}\,}
\newcommand{\ltish}{\protect\raisebox{-0.4ex}{$\,\stackrel{<}{\scriptstyle\sim}\,$}}
\newcommand{\gtish}{\protect\raisebox{-0.4ex}{$\,\stackrel{>}{\scriptstyle\sim}\,$}}

\newcommand{\ket}[1]{\mbox{$\left| #1 \right>$}}

\newcommand{\sub}[1]{\mbox{$_{\mbox{\tiny #1}}$}}
\newcommand{\diff}[1]{\mbox{\/d$#1$}}

\newcommand{\Ttwostar}{\mbox{$T_{2}^{\ast}$}}

\begin{document}
	
	\title{Probing Qubit Memory Errors at the Part-per-Million Level}
	
	\author{M.~A.~Sepiol}
	\author{A.~C.~Hughes}
	\author{J.~E.~Tarlton}
	\author{D.~P.~Nadlinger}
	\author{T.~G.~Ballance}
	\author{C.~J.~Ballance}
	\author{T.~P.~Harty}
	\author{A.~M.~Steane}
	\author{J.~F.~Goodwin}
	\email{joseph.goodwin@physics.ox.ac.uk}
	\author{D.~M.~Lucas}
	\address{Department of Physics, University of Oxford, Clarendon Laboratory, Parks Road, Oxford OX1 3PU, U.K.}
	
	\date{13 September 2019}
	
	\begin{abstract}
		Robust qubit memory is essential for quantum computing, both for near-term devices operating without error correction, and for the long-term goal of a fault-tolerant processor. We directly measure the memory error $\epsilon_m$ for a \Ca{43} trapped-ion qubit in the small-error regime and find $\epsilon_m<10^{-4}$ for storage times $t\ltish50\ms$. This exceeds gate or measurement times by three orders of magnitude. Using randomized benchmarking, at $t=1\ms$ we measure $\epsilon_m=1.2(7)\times10^{-6}$, around ten times smaller than that extrapolated from the \Ttwostar\ time, and limited by instability of the atomic clock reference used to benchmark the qubit. 
	\end{abstract}
	
	\maketitle
	
	A requirement for any physical realization of a quantum computer is the capability to preserve and to exploit the coherent behavior of its constituent qubits. During the course of a quantum computation, environmentally induced noise and imperfections in the control apparatus inevitably lead to a dephasing of qubit superpositions, which introduces memory errors $\epsilon_m$. While contributions to this error stemming from correlated noise can be suppressed using methods such as dynamical decoupling~\cite{Uhrig2007,Biercuk2009,Szwer2010}, quantum error correction (QEC) techniques~\cite{Shor1995,Steane1996} are necessary to counteract the remaining error. In a typical QEC circuit, ancilla qubits need to be prepared, entangled with logic qubits, and measured before appropriate feedback is applied to the logic qubits. Thus it is essential that the memory error remains below a correctable level at least for the duration of these operations, and preferably for even longer, to reduce the frequency with which ``idle'' qubits need to be corrected. Depending on the QEC methods used, and the architecture of the computer, the maximum correctable error can be as high as $\sim 10^{-2}$~\cite{Fowler2012b}, but a level of $10^{-4}$ is often taken as an important target for realistic overheads~\cite{Steane2007}. The longer $\epsilon_m$ remains correctable, the more flexibility there is in both the physical and logical design of the computer, and the greater the scope for increasing the connectivity of qubits (e.g., by moving the qubits around physically~\cite{Wineland1998}).
	
	For qubits with sufficiently long depolarization lifetimes ($T_1$), the memory error $\epsilon_m$ is determined by the relative stability of the qubit and reference oscillator frequencies~\cite{Ball2016}. The variation of $\epsilon_m$ with storage time $t$ depends on the spectrum of the noise processes that affect these frequencies. The benchmark almost universally used to quantify the memory performance is the coherence time \Ttwostar, the time constant for decay of qubit phase coherence when modeled by exponential decay $\exp\left(-t/\Ttwostar\right)$. The exponential decay model assumes that the spectrum of the frequency noise is white. \Ttwostar\ can be obtained by measuring the fringe contrast in Ramsey experiments as a function of the Ramsey delay $\tau_R$. For $\tau_R\,\ish\Ttwostar$, the memory error is much larger than qubit state preparation and measurement (SPAM) errors (typically $\epsilon\sub{SPAM}\gtish 10^{-3}$), and can therefore be measured easily. In such a method, information regarding the initial stages of decoherence---the regime relevant to quantum computing---has to be inferred by extrapolation, as in this regime $\epsilon_m \ltish \epsilon\sub{SPAM}$ so the large amount of data needed to measure $\epsilon_m$ leads to impractically long experiments. Consequently, any nonexponential structure to the coherence decay at short timescales remains undetected, leaving uncertainty about the true impact of memory errors on the computer's operation.
	
	Previous studies of qubit decoherence have measured \Ttwostar\ ranging from minutes to hours in large ensembles of trapped ions or nuclear spins~\cite{Bollinger1991,Fisk1995,Saeedi2013,Zhong2015}. For single physical qubits, a Ramsey $\Ttwostar\approx 50\s$ was measured for a \Ca{43} ion~\cite{Harty2014}, and a coherence time of $T_2\,\ish10\min$ was obtained by applying dynamical decoupling pulses to a \Yb{171} ion~\cite{Wang2017}. Only a single study~\cite{OMalley2015} has attempted to quantify the memory fidelity in the low-error regime of interest for quantum computing: in that work, a technique based on interleaved randomized benchmarking (IRB) was introduced to measure memory errors much smaller than $\epsilon\sub{SPAM}$, and was used in a superconducting qubit system to show that $\epsilon_m$ reached $\ish10^{-3}$ after a time equal to the typical duration of a single entangling gate. In the present Letter, we characterize the memory performance of a \Ca{43} trapped-ion hyperfine ``atomic clock'' qubit (frequency 3.2\GHz) both directly, using Ramsey experiments with short delays and high statistics, and indirectly, using the IRB method. 
	
	We use a microfabricated, planar surface-electrode ion trap that incorporates integrated microwave circuitry (resonators, waveguides, and coupling elements), and that is operated at room temperature. Its design, the details of which can be found in~\cite{Allcock2013}, allows for single- and two-qubit quantum logic gates to be driven by near-field microwave radiation instead of by lasers, eliminating photon-scattering errors and offering improved prospects for scalability~\cite{Ospelkaus2008,Ospelkaus2011,Harty2016}. For coherent manipulation of the qubit, we apply microwave pulses to electrodes which lie 75\um\ below the ion (see Supplemental Material \footnote{See Supplemental Material, which includes Ref.~\cite{Merkel2019}, for further details.}, \S{}A1). The experimental control and timing of the pulse sequences (with ns precision and ps jitter) is handled via ARTIQ~\cite{Bourdeauducq2016}.
	
	The main source of memory decoherence in trapped-ion hyperfine qubits typically stems from fluctuating magnetic fields. An established method for suppressing the effect of these fluctuations is to use a qubit based on an ``atomic clock'' transition whose frequency is independent of magnetic field to first order~\cite{Bollinger1991,Langer2005,Harty2014,Wang2017}. In this Letter, we use the intermediate-field \Ca{43} clock qubit formed by the $\ket{\downarrow}=\hfslev{4S}{1/2}{4,0}$ and $\ket{\uparrow}=\hfslev{4S}{1/2}{3,+1}$ ground-level hyperfine states~\footnote{Here we use the spectroscopic notation nL$_{\textrm{J}}^{\textrm{F}, \textrm{m}_{\textrm{F}}}$} at a static magnetic field of $B_0\approx146\G$~\cite{Harty2014}. We limit the effect of the qubit's second-order field dependence by stabilizing the field to within $\Delta B \ltish 1\mG$ of the field-independent point $B_0$ (see Supplemental Material [21], \S{}A2).
	
\begin{figure}
		\includegraphics[width=\onecolfig]{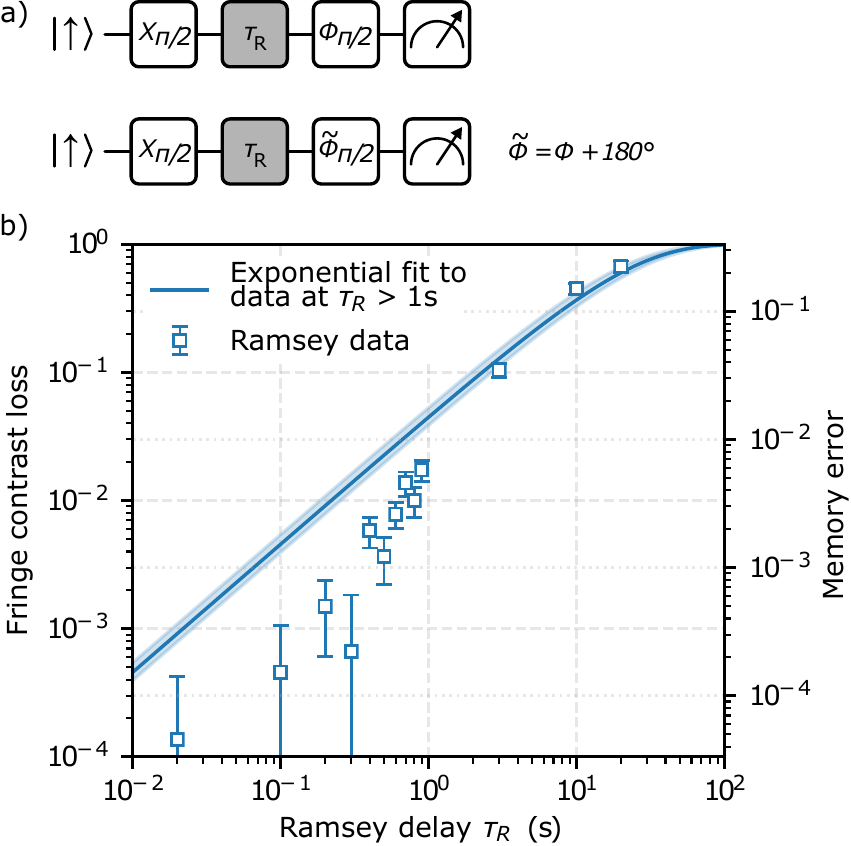}
		\caption{%
			Two-point Ramsey experiments for measuring the qubit memory error. %
			(a) Sequences used to measure the maximum and minimum of the Ramsey fringe. The phase difference between the final $\pi/2$ pulses remains fixed at $\tilde{\phi}-\phi=180\degree$ throughout the experiments.
			(b) SPAM-corrected loss of Ramsey fringe contrast. The line represents exponential contrast decay (fixed at 0 for $\tau_R=0$) fitted to the data with $\tau_R>1\s$, where the SPAM error is negligible compared with the contrast error; the fit gives $\Ttwostar=22(3)\s$. The right-hand ordinate gives the memory error, averaged over the Bloch sphere, associated with a given Ramsey contrast loss.
		}
		\label{F:Ramsey_exp}
	\end{figure}

	We first measure the memory error directly, using conventional Ramsey experiments in which the phase $\phi$ of the second $\pi/2$ pulse is varied relative to that of the first [Fig.\ \ref{F:Ramsey_exp}(a)]. We measure the Ramsey fringe contrast by fixing $\phi=\phi_0$ or $\phi=\phi_0+180\degree$, for fixed offset $\phi_0$, rather by than scanning $\phi$ and fitting the resulting fringe with a floated phase offset or scaling factor to compensate for slow phase drifts (as was done in prior work~\cite{Langer2005,Harty2014}). This ensures that our measurement is sensitive to phase drifts ($Z$ rotations) of the qubit relative to the microwave local oscillator, as would be the case during a quantum computation. The offset $\phi_0$ is calibrated before (but not during) experimental runs to compensate for any residual detuning offset of the microwaves (see Supplemental Material [21], \S{}A1).
	
	Since we aim to measure any loss in the fringe contrast (which is ideally 1) at a similar level to our $\epsilon\sub{SPAM}\approx 2\times 10^{-3}$, it is important to monitor drifts in $\epsilon\sub{SPAM}$, or systematic dependence of $\epsilon\sub{SPAM}$ on the Ramsey delay $\tau_R$. We follow each trial of the Ramsey sequence by two control sequences to measure $\epsilon\sub{SPAM}=\frac{1}{2}(\epsilon_\downarrow+\epsilon_\uparrow)$: the first consists of a delay $\tau_R$ between qubit \ket{\uparrow} state preparation and measurement; the second contains an additional $\pi$ pulse to prepare the \ket{\downarrow} state. We find that $\epsilon\sub{SPAM}$  shows negligible systematic variation for $\tau_R< 1\s$, and remains $\leq 10^{-2}$ for $\tau_R\leq 10\s$~\cite{JETThesis}. Alternating SPAM measurements with Ramsey experiments in this way allows us to capture any drifts in $\epsilon\sub{SPAM}$ with similar statistical uncertainty as for the Ramsey data. We subtract $\epsilon\sub{SPAM}$ from the measured contrast loss for each $\tau_R$.
	
	Results of this experiment, which required a week of continuous data acquisition, are shown in Fig.\ \ref{F:Ramsey_exp}(b). For a direct comparison with the IRB results (below), we can convert the loss in Ramsey fringe contrast to a memory error by scaling it by a factor of $\frac{1}{3}$: this takes into account that the error measured with IRB varies between 0 and $\frac{1}{2}$; and that in the IRB experiments the qubit state spends only $\frac{2}{3}$ of the time near the equator of the Bloch sphere, where it is sensitive to dephasing. The memory error is characterized down to a delay of $\tau_R=20\ms$, although for $\tau_R<200\ms$ uncertainty in $\epsilon\sub{SPAM}$ limits our knowledge of $\epsilon_m$ to an upper bound only. Collecting more data would reduce this uncertainty, but only with the square root of the acquisition time, making this approach impractical for exploring $\epsilon_m$ below the $10^{-4}$ level; it would also require systematic drifts in $\epsilon\sub{SPAM}$ to be $\ll 10^{-4}$. Figure \ref{F:Ramsey_exp}(b) also illustrates that assuming an exponential decay based on contrast measurements at long delays, as is customary in $\Ttwostar$ measurements, would here lead to a significant overestimate of the memory error for shorter delays.
	
	To circumvent the limitation imposed on memory error measurements by the SPAM error, we follow the approach introduced by O'Malley {\em et al.}~\cite{OMalley2015} which employs the technique of IRB. This method amplifies the memory error relative to the SPAM error by subjecting the qubit to $m$ periods $\tau$ of dephasing instead of to a single period, while ensuring that errors add incoherently by interleaving each delay $\tau$ with a Clifford gate $C_i$, sampled randomly from the full single-qubit Clifford group. We call the probability with which a given gate sequence of length $m$ produces the predicted final state the {\it sequence fidelity}, and for each $m$ we calculate the average sequence fidelity over $k=50$ distinct random sequences. The average sequence fidelity follows a decay $\frac{1}{2}(ap^{m}+b)$ with increasing $m$~\cite{Emerson2005,Knill2008}, where $p\in[0,1]$ is the {\it depolarizing parameter}, related to the average error rate by $\epsilon=\frac{1}{2}(1-p)$. Comparing the average sequence fidelity of IRB sequences with interleaved delays to that of ``reference'' RB sequences without delays allows us to isolate the average dephasing error associated with each delay, which we call the memory error $\epsilon_m = \frac{1}{2}(1-p_{\rm{IRB}}/p_{\rm{RB}})$. The depolarizing parameter of the reference sequences $p_{\rm{RB}}$ incorporates Clifford gate errors alone, while that of the interleaved sequences $p_{\rm{IRB}}$ also includes the memory errors from the delays. The parameter $a=1-2\epsilon\sub{SPAM}$ captures SPAM errors, which we ensure are equal in the reference and IRB sequences by adding a delay $m\tau$ after the final gate of the reference sequence. We fix $b=1$ as $\epsilon\sub{SPAM}\ll 1$. In this measurement of $\epsilon_m$, the precision attainable is no longer constrained by SPAM errors, but by the magnitude of the Clifford gate errors, while the accuracy attainable is limited by any systematic changes of the Clifford gate errors when delays are interleaved.
	
\begin{figure}
		\includegraphics[width=\onecolfig]{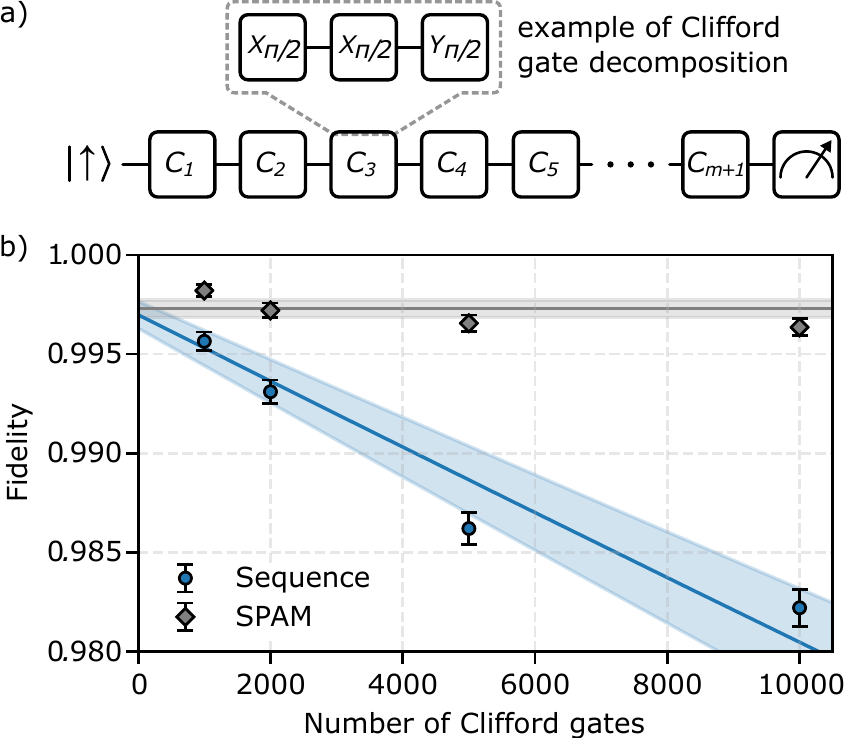}
		\caption{%
			``Standard'' randomized benchmarking of single-qubit gates.
			(a) Gate diagram of the sequence used, including an example decomposition of a Clifford gate $C_i$ into $\pi/2$ pulses. The qubit is prepared in $\ket{\uparrow}$ and then subjected to $m$ random gates from the Clifford set, followed by an ($m+1$)th gate that is chosen to rotate the qubit to one of its two basis states, selected with equal probability. Each RB sequence is alternated with a measurement of $\epsilon\sub{SPAM}$, for which the delay between state preparation and measurement is equal to the duration of the RB sequence; this allows us to check for any systematic dependence of $\epsilon\sub{SPAM}$ on sequence duration.
			(b) Measured sequence fidelities (blue circles) and SPAM fidelities (grey diamonds), as a function of sequence length. The fit to a decay $\frac{1}{2}(a p^m + 1)$ (blue line) gives an average Clifford gate error of $\epsilon_g=1.7(2)\times 10^{-6}$. The fit intercept is consistent with the mean measured SPAM error of $2.7(4)\times 10^{-3}$ (grey line). Error bars on each point are the standard error of the mean over $k=50$ sequence randomizations. Shaded regions represent the $1\sigma$ uncertainties of the fits.
		}
		\label{F:fidelity_rbm}
	\end{figure}

	We first characterize the Clifford gate errors {\em without} extra delays, using the ``standard RB'' (SRB) method~\cite{Emerson2005}, which involves applying random sequences of the form shown in Fig.\ \ref{F:fidelity_rbm}(a). The Cliffords are composed of $\pm X_{\pi/2}$ and $\pm Y_{\pi/2}$ rotations on the Bloch sphere, with an average of $3.50$ $\pi/2$ rotations per Clifford~\footnote{The minimal Clifford decomposition for $\{\pm X_{\pi/2},\pm Y_{\pi/2}\}$ physical gates and a `virtual' $Z_{\pi/2}$ leads to an average of 1.41 physical gates per Clifford. The longer decomposition used in this paper was chosen for historical reasons.}, separated by $12\us$ to allow the DDS source time to switch between pulse profiles. The duration of each pulse is set to be $\sim 10\us$ and periodically fine-tuned by optimizing the sequence fidelity for fixed $m$ (typically $m=2000$). Figure \ref{F:fidelity_rbm}(b) shows the measured sequence fidelity decay, yielding an average error per Clifford gate of $\epsilon_g=1.7(2)\times 10^{-6}$. A previous measurement of the gate error in this trap using the ``NIST RB'' method~\cite{Knill2008,Harty2014} (which used an average of two $\pi/2$ rotations per gate) gave $\epsilon_g=1.0(3)\times 10^{-6}$. Based on numerical modelling of known experimental imperfections~\cite{BoonePC}, the SRB and NIST RB methods are expected to yield a similar gate error; the $2\sigma$ discrepancy may be due to the larger number of physical $\pi/2$ rotations per gate used in the SRB experiment.

	\begin{figure*}
		\includegraphics[width=0.96\twocolfig]{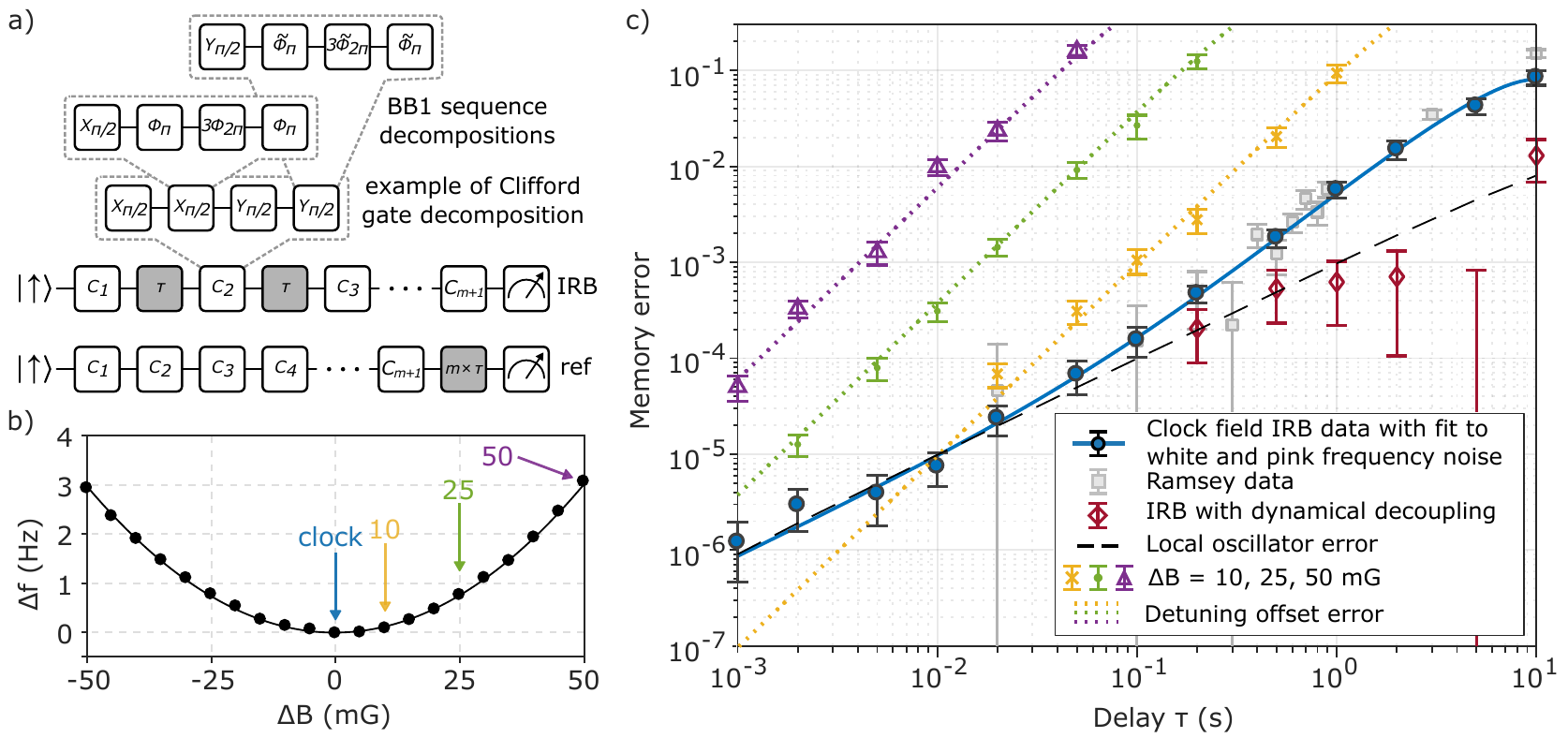}
		\caption{%
			Single-qubit memory errors $\epsilon_m$ measured by interleaved randomized benchmarking.
			(a) IRB sequence, with delays $\tau$ inserted between each Clifford gate $C_i$. An example decomposition of a Clifford gate $C_{i}$ into $+X_{\pi/2}$ and $+Y_{\pi/2}$ gates is shown. Also shown are the associated BB1 decompositions: each $X_{\pi/2}$ gate is followed by a sequence of rotations \{$\phi_{\pi}, (3\phi)_{2\pi}, \phi_{\pi}$\}; similarly for each $Y_{\pi/2}$ gate, but around an axis $\tilde{\phi}$. For the reference sequence (ref), an additional delay $m\tau$ is inserted after the last Clifford gate to keep the time between the qubit initialization and readout equal to that in the IRB sequence, thus minimizing any systematic differences in $\epsilon\sub{SPAM}$. 
			(b) Qubit frequency offset $\Delta f$ versus offset $\Delta B$ of the magnetic quantization field, relative to the field-independent ``clock'' point. 
			(c) Average memory error measured for interleaved delays $1\ms<\tau<10\s$. At each $\tau$, we choose $m$ to give a total sequence infidelity of $\ish 0.1$, which is much larger than the SPAM error. 
			The IRB data (blue circles) is consistent with the Ramsey results (grey squares, from Fig.\ \ref{F:Ramsey_exp}). We fit a decoherence model to the IRB data (blue line---see Supplemental Material [21], \S{}A3 for details), which contains white noise and $1/f$ noise with a low-frequency cutoff (this cutoff is consistent with the duration of the longest sequences).
			IRB sequences including dynamical decoupling (red diamonds) show that the error due to $1/f$ noise can be suppressed at long times using this technique. 
			The dashed black line shows the calculated error contribution from the local oscillator (atomic clock) to which the qubit phase is compared: this accounts for the measured memory error in the $\epsilon_m\ltish 10^{-4}$ regime. 
			Finally the yellow, green and purple data points show the error measured when the magnetic field is deliberately offset from the field-independent point by $\Delta B=(10,25,50)\mG$; note that the local oscillator was left at $\Delta f=0$, so that the observed error arises mostly from the corresponding qubit detuning $\Delta f\approx(0.12,0.76,3.0)\Hz$ (dotted lines).
		}
		\label{F:interleaved_rbm_exp}
	\end{figure*}

	Systematic variations of $\epsilon_g$ when the delays $\tau$ are interleaved can occur because the associated change in microwave duty cycle causes thermally induced shifts in the microwave power, and these variations would limit the accuracy of the IRB experiment if not prevented. Recalibrating the $\pi/2$-pulse time for each delay length $\tau$ by optimizing the sequence fidelity (as above) is impractical, as the interleaved delays significantly slow down the experiments. Instead, we make the $\pi/2$ rotations robust to pulse area changes by replacing them with BB1 composite gates~\cite{Wimperis1994}. This involves appending a $\{\phi_{\pi}, (3\phi)_{2\pi}, \phi_{\pi}\}$ sequence to each $\pi/2$ pulse [Fig.\ \ref{F:interleaved_rbm_exp}(a)], with $\phi$ appropriately chosen for every different $\pi/2$-pulse phase. Due to technical limitations of the DDS microwave source, implementing BB1 composite pulses required a Clifford gate decomposition into $+X_{\pi/2}$ and $+Y_{\pi/2}$ gates only, with an average of 3.58 $\pi/2$ rotations (i.e.\ 14.3 microwave pulses) per Clifford. The additional physical pulses necessary for the BB1 gates increase the average Clifford gate error to $\epsilon'_g=6(1)\times 10^{-6}$; no change in $\epsilon'_g$ was measured for a microwave pulse amplitude reduction of 1\%, much greater than that observed due to duty cycle effects.
	
	Results of the measurement of the qubit memory error $\epsilon_m$ using IRB with BB1 gates are shown in Fig.\ \ref{F:interleaved_rbm_exp}(c) (blue circles). During each data point, which requires up to two days of continuous acquisition, we measure the detuning between the qubit transition and the microwave source every four hours and correct for slow drifts; from these measurements, we estimate that the contribution to $\epsilon_m$ due to detuning errors is negligible~\cite{JETThesis}. Also plotted in Fig.\ \ref{F:interleaved_rbm_exp}(c) (grey squares) is the data from the Ramsey experiments (Fig.\ \ref{F:Ramsey_exp}). The memory error from both methods is consistent, but the superior sensitivity of the IRB approach enables characterization of $\epsilon_m$ for delays as short as $\tau=1\ms$, where, at $\epsilon_m=1.2(7)\times 10^{-6}$, it approaches the noise floor set by the BB1 Clifford gate error. We fit a decoherence model to the IRB data, which assumes white and pink ($1/f$) frequency noise. The memory error for $\tau\ltish 100\ms$ is consistent with that expected from independently measured $1/f^2$ phase noise (white frequency noise) on the rubidium atomic clock to which the microwave source is referenced (see Supplemental Material [21], \S{}A4). From decoherence measurements of a field-sensitive hyperfine transition, we estimate that contributions to the memory error from magnetic field noise are several orders of magnitude lower than our measured $\epsilon_m$ (see Supplemental Material [21], \S{}A2). The excess error for $\tau\gtish 100\ms$ may be due to other slow drifts, for example in the ac Zeeman shift ($\approx -6\Hz$) arising from the trap rf fields~\cite{TPHthesis}. We also took IRB data with a simple dynamical decoupling sequence (an $X_\pi$ pulse inserted every 100\ms\ during delays $\tau\ge 200\ms$), which suppressed the correlated ($1/f$) noise, reducing $\epsilon_m$ at longer delays to a level consistent with the atomic clock noise.
	
	An important consideration for a many-qubit processor based on the ``quantum CCD'' architecture~\cite{Wineland1998} will be inhomogeneity of the static magnetic field across the device, which will lead to departures $\Delta B$ of the field from the qubits' field-independent point $B_0$. To simulate the effect of field inhomogeneity, we took further IRB data with $\Delta B$ set to several offset values [Fig.\ \ref{F:interleaved_rbm_exp}(b) and \ref{F:interleaved_rbm_exp}(c)]. To obtain a ``worst case'' error, we did not adjust the microwave detuning for the known offset $\Delta f$; even so, for $\Delta B=50\mG$, we find $\epsilon_m < 10^{-4}$ at $\tau=1\ms$. This error is dominated by the effect of the (known) detuning offset, here $\frac{2}{3}\left[1-\cos^2(\pi\tau\Delta f)\right]\approx 6\times 10^{-5}$, which could be corrected for by calibrating the field across the processor.
	
	In conclusion, we have characterized the memory errors of a \Ca{43} trapped-ion hyperfine clock qubit to the $10^{-4}$ level by Ramsey measurement, and to the $10^{-6}$ level using interleaved randomized benchmarking. 
	The error is consistent with $1/f$ frequency noise at long timescales, contrary to the white noise model often assumed in $\Ttwostar$ measurements. The memory error remains below the $10^{-4}$ level relevant to QEC for up to 50\ms, which is around three orders of magnitude longer than the time required for entangling gates or qubit measurement~\cite{Ballance2016, Gaebler2016, Schafer2018, Crain2019}. At these sub-$10^{-4}$ error levels, the memory error is consistent with the independently-measured phase noise of the local oscillator, implying that a more stable reference clock would lead to improved performance.

	We thank J.~Emerson and K.~Boone for useful discussions, and M.~W.~Lucas for the loan of a time-to-amplitude converter. The ion trap was designed by D$^{3}$.T.C. Allcock. J.E.T.\ acknowledges funding from the Centre for Doctoral Training on Controlled Quantum Dynamics at Imperial College London. This work was supported by the U.S.\ Army Research Office (Ref. No. W911NF-14-1-0217) and the U.K.\ EPSRC ``Networked Quantum Information Technology'' Hub.


	\bibstyle{plain}


	%


	\pagebreak

	\setcounter{figure}{3}

    \section{Supplemental Material}

	\subsection{A1. Microwave source}

	The microwaves for the single-qubit gates are generated from a commercial direct digital synthesis (DDS) source~\footnote{Analog Devices AD9910/PCBZ DDS evaluation board}, whose $\ish400\MHz$ output is frequency-octupled, amplified and filtered, before being fed to one of the trap's microwave electrodes via a pair of solid-state switches (two switches are used for improved extinction during delays between gates). The DDS source is clocked at $990\MHz$ by a frequency synthesizer~\footnote{Rohde and Schwarz SMA 100A signal generator}, which is phase-locked to a rubidium atomic frequency standard~\footnote{Stanford Research Systems FS725 rubidium frequency standard}. To minimize pulse envelope distortions due to thermally-induced microwave power transients and variable duty cycles, the microwave line (up to the input of the on-chip resonator) is kept consistently warm by switching the source to a $\sim600\MHz$-detuned ``dummy'' signal of similar amplitude in the delays between pulses, which is generated by a separate microwave source and is introduced into the microwave drive network via one of the switches. This dummy signal causes an ac Zeeman shift on the qubit of $\sim 100\mHz$ during the delay periods, for which we compensate with a corresponding shift to the microwave oscillator frequency.

	The Ramsey experiments (Fig.\ 1) are very sensitive to the detuning between the microwave frequency and the qubit frequency for long delays $\tau_R$. We observe drifts in the detuning at the $\ish100\mHz$ level over periods of several days, which would cause the phase offset $\phi_0$ to be incorrect if not tracked. (A possible source of these slow drifts is variation of the ion position relative to the 38\MHz\ rf currents in the trap electrodes, which are observed to cause an ac Zeeman shift on the qubit transition of $-6\Hz$ at our typical operating parameters~\cite{TPHthesis_supp}.) To correct for these drifts, every 4 hours we map out the full Ramsey fringe by scanning $\phi$ over $360\degree$. As the resulting mHz-level frequency adjustments required are generally smaller than the resolution limit of the microwave source, we use the phase offset $\phi_0$ of the second $\pi/2$ pulse for fine tuning. To achieve longer-term fine frequency adjustments, we make Vernier-type adjustments by using the incommensurate minimum frequency steps of the DDS and clock synthesizer.\\

	\begin{figure}
		\includegraphics[trim={3.6cm 9.25cm 4.2cm 9.5cm}, clip, width=\onecolfig]{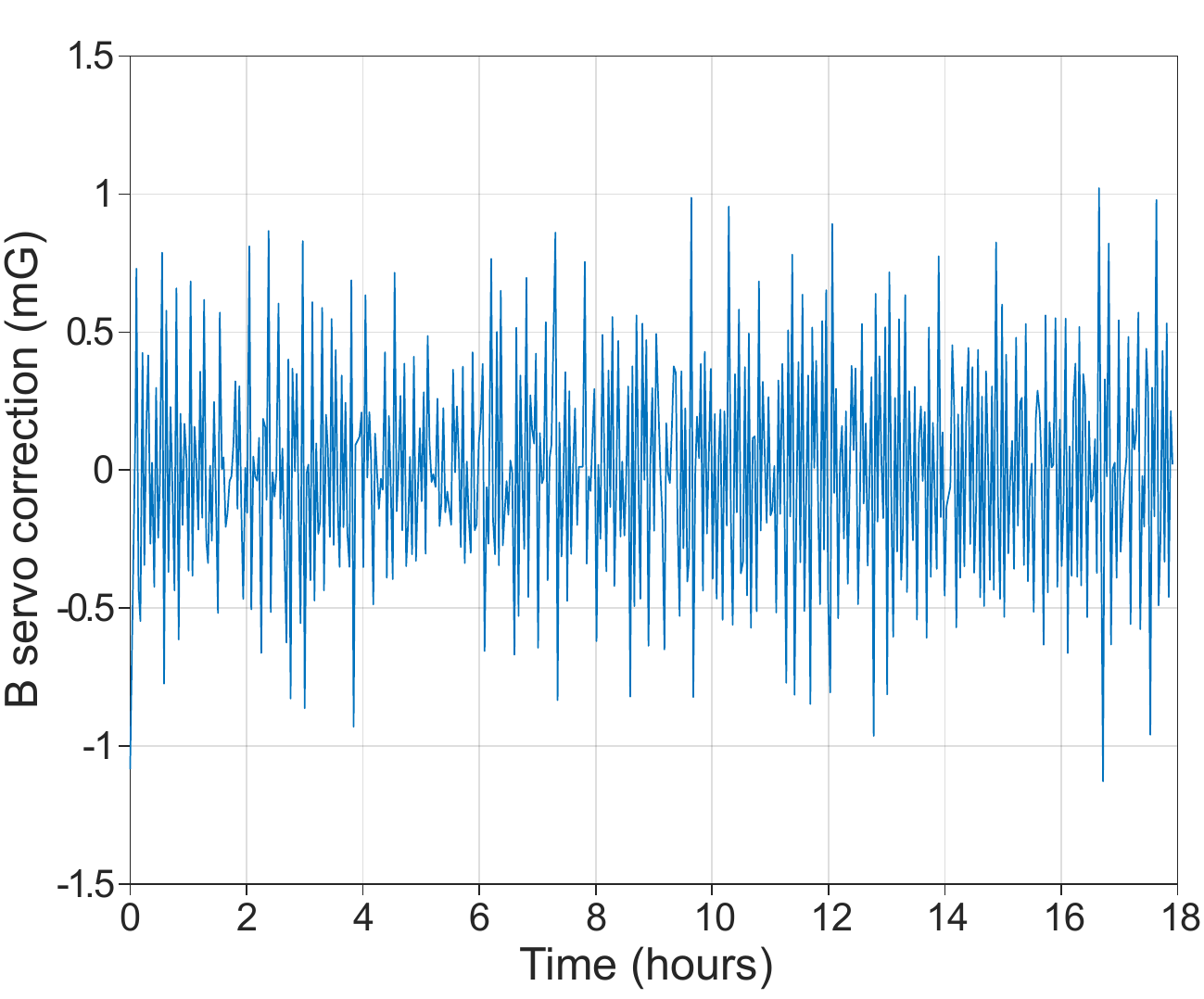}
		\caption{Time series of corrections made to the magnetic field after each application of the field servo routine, referenced to the frequency of the first-order field-sensitive $\hfslev{4S}{1/2}{4,+4}\leftrightarrow\hfslev{4S}{1/2}{3,+3}$ transition. Here we servo the field every $2\min$; the r.m.s.\ deviation of the corrections is $0.4\mG$ (c.f.\ the servo's statistical uncertainty of $\pm0.04\mG$).}
		\label{SF:servo_correct}
	\end{figure}

	\begin{figure*}
		a)\hfill~~~b)\hfill~\\ \vspace{-3ex}
		\includegraphics[width=\twocolfig]{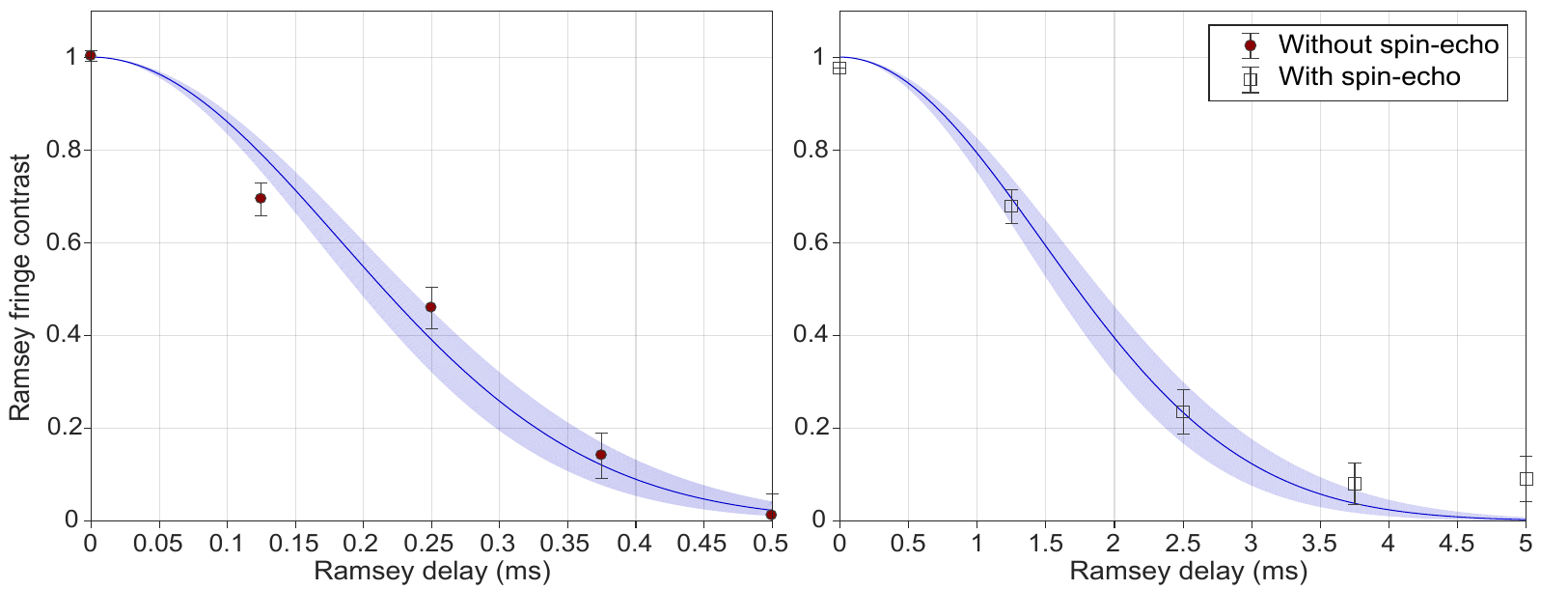}
		\caption{Ramsey contrast versus delay for the first-order field-sensitive $\hfslev{4S}{1/2}{4,+4}\leftrightarrow\hfslev{4S}{1/2}{3,+3}$ `stretch' transition, (a) without and (b) with the addition of a single spin-echo $\pi$-pulse. Note the difference in horizontal scale. The data are well-described by the fitted Gaussian decays. The spin-echo data implies a coherence time of $T_2=2.1(2)\ms$, attributable to magnetic field noise with a $1/f$ frequency dependence and r.m.s amplitude of $47(4)\uG$ over the duration of the experiment. The reduced coherence time of $T_2^*=0.26(2)\ms$ without spin-echo is consistent with $\sim0.3\mG$ field offsets constant over the duration of the experiment.}
		\label{SF:stretch_deco}
	\end{figure*}

	\subsection{A2. Magnetic field stabilization}

	The atomic clock qubit frequency $f\sub{clk}$ is insensitive to magnetic field to first order at the field-independent point, $B_0\approx 146\G$, with a second-order dependence $\diff{^2 f\sub{clk}}/\diff{B^2} = 2.42\mHz/\mG^2$. For optimum memory performance of the qubit, we aim to keep the static field $B$ within $\Delta B \ltish 1\mG$ of the field-independent point, using four automated control mechanisms, each acting on the $\ish135\amp$ current applied to the pair of coils used to generate the field. The first is a feedback system~\cite{Merkel2019_supp} that suppresses coil current variations below the loop's bandwidth of $3\kHz$, using a fluxgate current sensor for measuring the fluctuations, and a transistor-controlled current shunt, which diverts a small fraction of the current from the coils. The second is a feedforward system acting on the coil current that helps suppress the effect of field noise synchronous with the 50\Hz\ mains electricity cycle (from both the coil current supply and laboratory environment); based on a series of iterative calibration measurements on the ion, we reduce field noise at $50\Hz$ and the first 7 harmonics of this frequency by a factor of $\ish4$, to $\ish0.15\mG$ r.m.s. The third is a feedback system for correcting slower, uncorrelated drifts in the laboratory field, using readings taken every 2\s\ from a magnetic field fluxgate sensor placed near the ion trap. Finally, each experimental run is interrupted at approximately equal time intervals (typically every 2\min) to correct for any field drifts that have not been cancelled by the feedforward from the fluxgate sensor, via a servo routine that uses the ion itself as a probe, by measuring the frequency of the $\hfslev{4S}{1/2}{4,+4}\leftrightarrow\hfslev{4S}{1/2}{3,+3}$ transition which is first-order field-sensitive. These small residual offsets may be due to sources relatively near the trap, which give unequal field magnitude at the sensor and at the ion. The statistical precision of this servo is $\pm0.04\mG$. This final lock to the atomic structure of the ion provides an absolute field reference, which also allows us to maintain a desired field offset $\Delta B$ with similar precision.

	Slowly varying residual offsets of the field from the target value with all control mechanisms operational can be estimated from the corrections made after each application of the servo routine, as shown in Fig.\ \ref{SF:servo_correct}. Applying the routine every 2\min\ ensures the field remains within $\sim\pm1\mG$ of the target value at all times, with an r.m.s.\ deviation of $0.4\mG$.

	The $\hfslev{4S}{1/2}{4,+4}\leftrightarrow\hfslev{4S}{1/2}{3,+3}$ transition also provides a useful probe of field noise at shorter timescales, which we measure by performing Ramsey experiments and observing the decay of fringe contrast versus Ramsey delay. In Fig.\ \ref{SF:stretch_deco}(a) we show the results of such a measurement, which exhibit a Gaussian decay of contrast $\exp{[-(t/T_2^*)^2]}$ with fitted $T_2^*=0.26(2)\ms$. We also perform Ramsey experiments with the addition of a single `spin-echo' $\pi$-pulse halfway through the delay, which nulls the effect of a small static offset of the field during the Ramsey sequence. The results of the spin-echo experiment are shown in Fig.\ \ref{SF:stretch_deco}(b); the longer coherence time of $T_2=2.1(2)\ms$ can be associated with the residual fast variations of the field; from the sensitivity of the stretch transition $\diff{f\sub{str}}/\diff{B} = -2.36\MHz/\G$ we calculate r.m.s.\ magnetic field noise of $47(4)\uG$, while the Gaussian decay of contrast implies a field noise spectral density $S_B(f)\propto1/f$. The reduced coherence time measured without the echo pulse can be accounted for by a static frequency offset of $640\Hz$, i.e.\ a field offset of $0.27\mG$, which is well within the expected deviation of the servo.

	These measurements at the millisecond timescale allow us to put some constraints on the expected effect of magnetic field noise on the clock qubit memory errors at similar timescales. From the servo correction data (Fig.\ \ref{SF:servo_correct}), we can assume a typical field offset of $\Delta B = 0.4\mG$ during our experiments, at which the field sensitivity of the clock qubit is:
	\[
	\left(\frac{\diff{f\sub{clk}}}{\diff{B}}\right)_{\Delta B} \approx 0.97\Hz/\G \approx 0.39\times10^{-6}\left(\frac{\diff{f\sub{str}}}{\diff{B}}\right).
	\]
	\noindent The angular extent of dephasing of the clock qubit from the microwave oscillator after a given delay will be reduced by this factor relative to a qubit encoded on the stretch transition. At $t=T_2=2.1\ms$, when the spin-echo contrast was reduced to $1/\mathrm{e}$, the r.m.s.\ angular deviation of the dephased stretch qubit Bloch vector is expected to be $\Delta\phi\sub{str}\approx1.2\rad$; we would therefore expect a clock qubit dephasing angle of $\Delta\phi\sub{clk}\approx0.47\urad$. The dephasing of the clock qubit after $2\ms$ is thus predicted to introduce memory errors $\epsilon_m\ish 10^{-13}$. Assuming the $1/f$ frequency dependence implied by the stretch transition Ramsey experiments extends to lower frequencies, we therefore predict that the contribution of field noise to memory error will mirror that of the pink frequency noise contribution we observed at long timescales, but with $\ish 10^6$ times lower magnitude. In fact, in this system, the residual magnetic field noise is unlikely to limit performance until $\Delta B \gg100\mG$.

	\section{A3. Decoherence model}

	\vspace{-1ex}
	We fit the IRB data to a model including white and pink ($1/f$) frequency noise, where the total noise power spectral density (PSD) is given by
	\[
	S(f)=S_W + \frac{S_P}{f}.
	\]

	\noindent The memory error expected from such a noise spectrum is given by \cite{OMalley2015_supp}
	\[
	\epsilon_m(\tau)=\frac{\pi}{3}\left(\frac{S_W\tau}{2}+S_P\tau^2\textrm{ln}\left(\frac{0.4}{f_c\tau}\right)\right),
	\]
	where $S_W=0.0016(3)$ Hz, $S_P=0.0014(3)$ Hz$^2$ and $f_c=0.025(3)$~Hz are the fitted parameters. The low-frequency cut-off $f_c$ of the $1/f$ noise is consistent with the inverse of the total duration of the longest IRB sequences (40 seconds).

	\subsection{A4. Local oscillator phase noise characterization}

	\vspace{-1ex}
	We use two microwave (mw) sources in the apparatus, one (as described in \S{}A1) to drive mw pulses on the qubit transition, the other (which is similarly constructed) for qubit state preparation and readout pulses on other hyperfine transitions~\cite{Harty2014_supp}. For all the experiments described in the paper, the two sources share a common rubidium clock reference and synthesizer. For the purposes of characterizing the phase noise in the mw system, these two nominally similar mw sources were set to the same frequency and their outputs were combined on a mixer to make a homodyne measurement. By comparing the phase noise using (i) the same Rb clock and synthesizer, (ii) the same Rb clock and two identical synthesizers, and (iii) two independent Rb clocks driving two identical synthesizers, we found that the Rb clocks were the dominant source of phase noise. The power spectral density of the phase noise was measured on an oscilloscope and was found to have an approximately $1/f^2$ dependence (i.e.\ consistent with white frequency noise) for $1\Hz < f < 10\kHz$. Drift at longer timescales was measured with a digital voltmeter and found to be consistent with white frequency noise of the same spectral density ($S^\textrm{\tiny{Rb}}_W(f)=0.0019(2)$ Hz for $1\mHz < f < 0.2\Hz$. This measured noise was used to calculate the ``local oscillator error'' curve in Fig.\ 3(c), with no free parameters, and is consistent with the measured IRB memory error for $\tau\ltish 100\ms$. It is also consistent with the memory error measured using IRB with dynamical decoupling for $\tau\gtish 100\ms$.

	\bibstyle{plain}


	%

\end{document}